\documentclass[reprint,showpacs,amsmath,amssymb,aps,pra]{revtex4-1}

\usepackage{amsfonts}
\usepackage{amsmath}
\usepackage{amssymb}
\usepackage{graphicx}
\usepackage{bm}
\usepackage{hyperref}
\usepackage{subfigure}

\begin{document}
\title{Quantum Coherent Absorption of Squeezed Light}

\author{Ali \"{U}. C. Hardal$^1$}
\email{alhar@fotonik.dtu.dk}
\author{Martijn Wubs$^{1,2}$}
\affiliation{$^1$ Department of Photonics Engineering, Technical University of Denmark, \O rsteds Plads 343, DK-2800 Kgs. Lyngby, Denmark}
\affiliation{$^2$ Center for Nanostructured Graphene, Technical University of Denmark, DK-2800 Kongens
	Lyngby, Denmark}

\begin{abstract}
We investigate coherent perfect absorption (CPA) in quantum optics, in particular when pairs of squeezed coherent states of light are superposed on an absorbing beam splitter. First, by employing quantum optical input-output relations, we derive the absorption coefficients for quantum coherence and for intensity, and reveal how these will differ for squeezed states. Secondly, we present the remarkable properties of a CPA-gate: two identical but otherwise arbitrary incoming squeezed coherent states can be completely stripped off their coherence, producing a pure entangled squeezed vacuum state that with its finite intensity escapes from an otherwise perfect absorber. Importantly, this output state of light is not entangled with the absorbing beam splitter by which it was produced. Its loss-enabled functionality makes the CPA gate an interesting new tool for continuous-variable quantum state preparation.
\end{abstract}
\maketitle

\section{Introduction}
Coherent perfect absorption (CPA) of light~\cite{chong2010coherent} is an interference-assisted absorption process that in its simplest form can take place when two coherent beams impinge on the opposite sides of an absorbing beam splitter. With input light from only one side, some light would leave the beam splitter, but no light emerges if there is equal input from both sides. While scattering 
theory can provide a  rigorous mathematical description, in essence, the reflected part of one of the incident beams interferes destructively with the transmitted part of the other (and vice versa), forming an artificial trap for the light that is subsequently dissipated~\cite{chong2010coherent}.
So unlike the usual incoherent absorption, the coherent absorption of light is an emergent property~\cite{holland2014complexity} 
that arises  
from a specific interplay of interference and dissipation~\cite{chong2010coherent,baranov2017coherent}. 

Many realizations of CPA  have been proposed, including homogeneously broadened two-level systems~\cite{longhi2011coherent}, epsilon-near-zero metamaterials~\cite{feng2012coherent}, graphene~\cite{Liu:2014a}, and heterogeneous metal-dielectric composite layers~\cite{dutta2012controllable}.  CPA has been successfully demonstrated in many setups, for example in a silicon cavity with two counterpropagating waves~\cite{wan2011time}, using a pair of resonators coupled to a transmission line~\cite{sun2014experimental}, and using graphene to observe CPA of optical~\cite{Pirruccio:2013a} and of  terahertz radiation~\cite{Kakenov:2016a}.  Achieving CPA under single-beam illumination with perfect magnetic conductor surfaces has also been reported~\cite{li2014equivalent}. A recent study revealed that CPA of light can be used as an unconventional tool to strongly couple light to surface plasmons in nanoscale metallic systems~\cite{noh2012perfect}. The fields in which CPA may play a central role include but are not limited to photodetection~\cite{konstantatos2010nanostructured,knight2011photodetection}, sensing~\cite{liu2010infrared}, photovoltaics~\cite{luque2011handbook} and cloaking~\cite{salisbury1952absorbent,vinoy1996radar}. For further details and examples of CPA realizations we refer to the excellent recent review by Baranov et al.~\cite{baranov2017coherent}.

Investigations of CPA in the quantum regime (also known as Quantum CPA~\cite{baranov2017coherent}) have only recently begun, notably with entangled few-photon input states~\cite{Huang:2014a,Roger2015coherent,Roger:2016a,Altuzarra:2017a}. Here instead, we consider squeezed coherent states of light~\cite{walls1983squeezed} and report the effects of quadrature squeezing on the absorption profile of the system, and vice versa the effects of the absorbing beam splitter on the generated output states of light. To the best of our knowledge, this important class of continuous-variable quantum states of light has not yet been considered in the context of CPA.

Squeezed states  of light have no classical analogues. They reduce noise in one field quadrature at the expense of larger noise in the other, such that Heisenberg's uncertainty relation for their product holds. First realized decades ago~\cite{andersen201630}, squeezed states of light continue to attract attention, 
mainly due to their indispensable roles in quantum information, communication and optics protocols. In particular, squeezed states are key for quantum teleportation~\cite{milburn1999quantum,van2000multipartite,yonezawa2004demonstration}, quantum key distribution~\cite{cerf2001quantum}, quantum metrology~\cite{anisimov2010quantum}, quantum cryptography~\cite{hillery2000quantum}, quantum dense coding~\cite{ban1999quantum}, quantum dialogue protocols~\cite{zhou2017new}, quantum laser pointers~\cite{treps2003quantum}, 
and quantum memories~\cite{jensen2011quantum,appel2008quantum}. They can be used to distinguish quantum states by enhancing quantum interference~\cite{tombesi1987generation} and for robust electromagnetically induced transparency~\cite{akamatsu2004electromagnetically}. They are used to increase the sensitivity of gravitational wave detectors~\cite{aasi2013enhanced} as well. 
While usually produced in macroscopic setups~\cite{andersen201630}, squeezed light may also be produced by (pairs of) individual emitters in optical nanostructures~\cite{MartinCano:2017a}. Recently, squeezed vacuum was proposed to engineer interactions between electric dipoles~\cite{zeytinouglu2017engineering}. 

In this contribution, we  distinguish between the usual absorption of intensity and that of quantum coherence, the latter measured as the coherent degree of freedom in Glauber's sense~\cite{glauber1963quantum}. The latter measure is conveniently chosen such that the two measures are equivalent for coherent states,  but for squeezed coherent input states we show that they differ.
For the latter case, we will show that under the CPA conditions, a one- and two-mode combined squeezed vacuum state~\cite{hong1990squeezed,abdalla1992statistical1,abdalla1992statistical,yeoman1993two} is produced, a finding that does not rely on our definition of coherence. Meanwhile all coherence of the  input states is transferred to the internal modes of the beam splitter. Importantly, we will show that in the output state the light is not entangled with the lossy beam splitter. These and  further intriguing properties 
may make the lossy CPA beam  splitter  a useful element in continuous-variable protocols.    

The interference of waves depends on their statistical properties. With quantum states of light as inputs, it is then natural to look for  connections between CPA and quantum statistics. Here we express the coherent absorption coefficient in terms of the fidelity~\cite{uhlmann1992metric} between two incoming coherent states. The requirement for perfect absorption of coherence becomes the complete indistinguishability of the incoming fields. This requirement also holds for squeezed coherent states of light as input. 

This paper is organized as follows. In Sec.~\ref{sec:results}, we revisit CPA from first principles and distinguish between the absorption of quantum coherence and intensity, and we identify a statistical connection between the coherent absorption coefficient and the fidelity of the input states. In Sec.~\ref{sec:CPA_sq}, we present coherent absorption of squeezed coherent input states. In Sec.~\ref{sec:CPA_qsp} we derive and discuss the remarkable quantum state transformation that is performed by the CPA beam splitter, and we conclude in Sec.~\ref{sec:conc}.

\section{CPA revisited: coherent states}\label{sec:results}
\subsection{Basic setup and key concepts}\label{sec:setup_and_concepts}
\begin{figure}[t!]
	\centering
	\begin{center}
		\includegraphics[width=6cm]{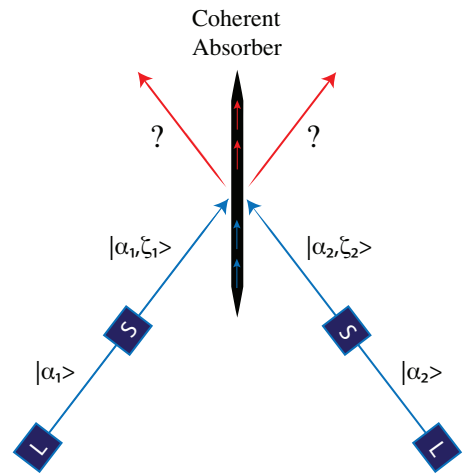}
	\end{center}
	\caption{ A sketch of the model system: two quantum states of light are superposed on an absorbing beam splitter. We consider squeezed coherent states that for example can be prepared by sending coherent laser (L) light through a crystal that produces squeezing (S). Squeezing affects the coherent absorption. In turn the CPA beam splitter gives rise to dissipation-enabled preparation of pure entangled two-mode output squeezed vacuum states, see main text.   
	}
	\label{fig:fig1}
\end{figure}

Let us consider a lossy beam-splitter in free space which superposes two incident quantized modes of light and creates two outgoing modes as shown in Fig.~\ref{fig:fig1}. The incident modes are described by the discrete annihilation operators $\hat{a}_{1}$ and $\hat{a}_{2}$. The field operators $\hat{b}_{1}$ and $\hat{b}_{2}$ of the outgoing modes are, then, given by the relations~\cite{PhysRevA.52.4823,PhysRevLett.77.1739,PhysRevA.57.2134,Knoell:1999a}
\begin{subequations}\label{eq:in_out}
\begin{eqnarray}
\hat{b}_{1}&=&t\hat{a}_{1}+r\hat{a}_{2}+\hat{L}_{1},\\
\hat{b}_{2}&=&r\hat{a}_{1} + t\hat{a}_{2}+\hat{L}_{2},
\end{eqnarray}
\end{subequations}
where $t$ and $r$ are the beam-splitter's transmission and reflection amplitudes and $\hat{L}_{1}$ and $\hat{L}_{2}$ describe Langevin-type noise operators corresponding to device modes of the beam splitter in which the light absorption takes place. We discuss this quantum noise in Sec.~\ref{sec:CPA_qsp}. We adopted the discrete-mode representation of the quantized fields, but there would be ways to generalize this   to full continuum~\cite{TualleBrouri:2009a}. 

In the lossy beam splitter, the light will typically loose part of its coherence and also part of its intensity, both due to a combination of destructive interference and dissipation. It will be enlightening to distinguish between the absorption of intensity and of coherent amplitudes. 

We define the total intensities of incoming and outgoing fields in the standard way as 
\begin{equation}
\mathcal{I}_{\text{in}} \equiv \langle\hat{a}^{\dagger}_{1}\hat{a}_{1}\rangle+\langle\hat{a}^{\dagger}_{2}\hat{a}_{2}\rangle \quad \mbox{and} \quad\mathcal{I}_{\text{out}} \equiv\langle\hat{b}^{\dagger}_{1}\hat{b}_{1}\rangle+\langle\hat{b}^{\dagger}_{2}\hat{b}_{2}\rangle,
\end{equation}
and the lost intensity as $\Delta\mathcal{I} \equiv \mathcal{I}_{\text{in}} - \mathcal{I}_{\text{out}}$. The coefficient of absorption of intensity is  
\begin{equation}\label{eq:absorbedintensity}
\mathcal{A}_{\text{coh}}^{\mathcal{I}} \equiv 1 - \mathcal{I}_{\text{out}}/\mathcal{I}_{\text{in}},
\end{equation}
being the fraction of intensity that gets lost. 
Analogously, we choose to quantify the input coherent amplitudes through the quantity
\begin{equation}\label{eq:input_coherence}
\mathcal{C}_{\text{in}} \equiv |\langle\hat{a}_{1}\rangle|^2+|\langle\hat{a}_{2}\rangle|^2
\end{equation}
i.e. as the sum of the absolute values squared of the expectation values $\langle\ldots\rangle$, the latter taken with respect to the initial quantum state $|\psi\rangle_{\text{in}}$. This state describes both the  quantum state of light in both arms and of the internal states of the beam splitter. We give a rationale for this measure of coherent amplitudes below. Correspondingly, we quantify the output coherent amplitude,
$\mathcal{C}_{\text{out}} \equiv |\langle\hat{b}_{1}\rangle|^2+|\langle\hat{b}_{2}\rangle|^2$,  
and the net loss in  coherent amplitudes as $\Delta \mathcal{C}\equiv \mathcal{C}_{\text{in}}-\mathcal{C}_{\text{out}}$. We define the coefficient of absorption of  coherent amplitudes as
\begin{equation}\label{eq:coh_ab_c0}
\mathcal{A}_{\text{coh}}^{\mathcal{C}} \equiv \Delta \mathcal{C}/\mathcal{C}_{\text{in}}= 1-\mathcal{C}_{\text{out}}/\mathcal{C}_{\text{in}}.
\end{equation}
This definition is state independent and well-defined for systems for which the input-output relations~\ref{eq:in_out} hold and the input coherence is non-vanishing. 
We will be especially interested in two specific situations: Coherent perfect absorption, corresponding to $\mathcal{A}_{\text{coh}}^{\mathcal{I}} \equiv 1$ (no output intensity), as opposed to perfect absorption of coherence, or $\mathcal{A}_{\text{coh}}^{\mathcal{C}} \equiv 1$ (vanishing output coherent amplitudes).

Let us elucidate our measure of coherent amplitudes~(\ref{eq:coh_ab_c0}), and let us first state what it is not.  It does not measure quantum correlations between different optical ports. Such  correlations of the form $\langle \hat{a}_{1}^{\dagger}\hat{a}_{2}\rangle$
are already encoded in the coefficients $\mathcal{C}_{\rm in,out}$ through the relations~(\ref{eq:in_out}) of our quantum optical input-output formalism. Secondly, it does not measure the most general case of quantum coherent resources introduced by arbitrary quantum states~\cite{streltsov2017colloquium}. Such generality is not needed here since we focus on coherent and squeezed coherent states.

So what does Eq.~(\ref{eq:coh_ab_c0}) measure? Throughout this paper, we consider coherent states $|\alpha\rangle$ as introduced by Glauber~\cite{glauber1963quantum}, and quantum states that can directly be defined in terms of them. The well known coherent states are eigenstates of the annhilation operator $ a|\alpha\rangle = \alpha|\alpha\rangle$, with the eigenvalue $\alpha$ being the complex coherent amplitude.
Therefore, the magnitude of the coherent amplitude of the \textit{fully coherent states} $|\alpha\rangle$~\cite{glauber1963quantum,klauder2006fundamentals} and of quantum states that can directly be obtained through them can be quantified by the absolute expected value $|\langle \hat{a}\rangle|$ of the corresponding bosonic annihilation operator $\hat{a}$~\cite{klauder2006fundamentals}. We intentionally choose to measure the square of $|\langle \hat{a}\rangle|$,  so that for coherent states our measure  will numerically coincide with that of intensity (see Sec.~\ref{sec:2B}). This  enables us to witness the breakdown of this equality in the case of squeezed states of light and thus, will provide a genuine new perspective to  quantum coherent perfect absorption.

Written out in the photon-number state basis, the average intensity only depends on populations (diagonal elements) of the density matrix describing the state of light:  $\langle \hat{a}^{\dagger}\hat{a}\rangle=\sum_{n}\rho_{nn}n$. By contrast, the expectation value of the annihilation operator only depends on off-diagonal matrix elements (also known as quantum coherences~\cite{streltsov2017colloquium}) of the density matrix:    
$\langle \hat{a}\rangle=\sum_{n}\rho_{nn-1}\sqrt{n}$. In particular, it only depends on the one-photon coherences, being the coherences between states that differ in photon number by one. It follows that the intensity absorption coefficient $\mathcal{A}_{\text{coh}}^{\mathcal{I}} $ of Eq.~(\ref{eq:absorbedintensity})  
only depends on the photon-number populations of the density matrix, whereas the absorption coefficient of coherent amplitudes $\mathcal{A}_{\text{coh}}^{\mathcal{C}}$ in Eq.~\ref{eq:coh_ab_c0}  only depends on its 1-photon coherences.  In the following for simplicity we refer to $\mathcal{A}_{\text{coh}}^{\mathcal{C}}$ as the absorption of (quantum) coherence. The coherences are bounded from above  by the average photon number as described by the inequality $|\langle \hat{a}\rangle|^2\leq\langle \hat{a}^{\dagger}\hat{a}\rangle$, which follows from 
the generalized Cauchy-Schwarz inequality~\cite{klauder2006fundamentals}.
\subsection{Coherent absorption of coherent states}\label{sec:2B}

Throughout this paper we make the common assumption that initially the internal device modes of the beam splitter are in their ground states, denoted as $|\rangle_{\text{BS}}$. 
For the incident optical modes, let us first consider that both are prepared in coherent states, also known as displaced vacuum states~\cite{scully1999quantum}. A coherent state (for example of  mode~1) is defined as $|\alpha\rangle_{1}=\hat{D}_{1}(\alpha)|\rangle$ 
in terms of the optical vacuum state $|\rangle$ and the displacement operator 
\begin{equation}\label{eq:displacementoperator}
\hat{D}_{1}(\alpha)\equiv\exp{(-|\alpha|^2/2)}\exp{(\alpha\hat{a}_{1}^{\dagger})}\exp{(-\alpha^*\hat{a}_{1})}.
\end{equation}
The total input state can then be written as  
$|\psi\rangle_{\text{in}}=|\alpha
\rangle_{1}\otimes|\beta
\rangle_{2}
\otimes|\rangle_{\text{BS}}$ with $\alpha=|\alpha|\exp{(i\theta_1)}$ and $\beta=|\beta|\exp{(i\theta_2)}$. 
For the input coherence of these coherent states we then find 
\begin{equation}
\mathcal{C}_{\text{in}}=|\alpha|^2+|\beta|^2,
\end{equation}
and for the output coherence 
\begin{eqnarray}
\nonumber \mathcal{C}_{\text{out}}
&=&(|\alpha|^2+|\beta|^2)(|t|^2+|r|^2)\\
&+&2\cos{(\theta)}|\alpha||\beta|(tr^{*}+rt^{*}),
\end{eqnarray}
where we used the input-output relations~\ref{eq:in_out} and  defined $\theta \equiv\theta_2-\theta_1$ as the phase difference between the coherent states. The coefficient of absorption of quantum coherence then reads
\begin{equation}\label{eq:coh_ab_c1}
\mathcal{A}_{\text{coh}}^{\mathcal{C}}=1-\bigg[|t|^2+|r|^2+\frac{2|\alpha||\beta|(tr{^*}+rt^{*})}{|\alpha|^2+|\beta|^2}\cos{(\theta)}\bigg],
\end{equation}
which reduces to the conventional expression for incoherent absorption $\mathcal{A}=1-(|t|^2+|r|^2)$ when replacing $\cos{(\theta)}$ by its average value of zero. 

For  two  coherent incident states, the  quantum coherence lost is in fact equal to the lost  intensity, because the identities  $\mathcal{C}_{\text{in}} = \mathcal{I}_{\text{in}}$ and $\mathcal{C}_{\text{out}} = \mathcal{I}_{\text{out}}$ hold in that case. This then  immediately implies $\Delta\mathcal{C}=\Delta\mathcal{I}$ and $\mathcal{A}_{\text{coh}}^{\mathcal{C}}=\mathcal{A}_{\text{coh}}^{\mathcal{I}}$. At this point it may seem pedantic that we first distinguished between these two coefficients, but as we shall see in Sec.~\ref{sec:CPA_sq}, this equality is not true for incident squeezed states of light.
\subsection{Fidelity and CPA}\label{sec:CPA_fidelity}
Next, we relate the expression~(\ref{eq:coh_ab_c1}) for coherent absorption of coherent states to  their quantum fidelity. As our starting point we recall that the inner product of  any pair of coherent states $|\alpha\rangle$ and $|\beta\rangle$  is~\cite{scully1999quantum}
\begin{equation}\label{eq:not_orthogonal}
\langle\alpha|\beta\rangle=e^{-\frac{1}{2}(|\alpha|^2+|\beta|^2)+\alpha\beta^{*}},
\end{equation}
which implies they are never orthogonal. It follows that
\begin{subequations}
\begin{eqnarray}
\alpha\beta^{*}&=&|\alpha||\beta|e^{-i\theta}=\frac{1}{2}(|\alpha|^2+|\beta|^2)+\ln{\langle\alpha|\beta\rangle},\\
\beta\alpha^{*}&=&|\alpha||\beta|e^{i\theta}=\frac{1}{2}(|\alpha|^2+|\beta|^2)+\ln{\langle\beta|\alpha\rangle}.
\end{eqnarray}
\end{subequations}
Hence, we obtain
\begin{eqnarray}
\nonumber2|\alpha||\beta|\cos{(\theta)}&=&|\alpha|^2+|\beta|^2+\ln{|\langle\alpha|\beta\rangle|^2}\\
&=&|\alpha|^2+|\beta|^2+\ln{F(\rho_{\alpha},\rho_{\beta})},
\end{eqnarray}
where $F(\rho_{\alpha},\rho_{\beta})$ is the Ulhmann's fidelity~\cite{uhlmann1992metric} with $\rho_{\alpha} \equiv|\alpha\rangle\langle\alpha|$ and $\rho_{\beta} \equiv|\beta\rangle\langle\beta|$. Written in terms of the fidelity, the coherent absorption coefficient~(\ref{eq:coh_ab_c1}) reads
\begin{equation}\label{eq:coh_ab_c2}
\mathcal{A}_{\text{coh}}^{\mathcal{C, I}}=1-\bigg[|t+r|^2+\frac{(tr^{*}+rt^{*})}{|\alpha|^2+|\beta|^2}\ln{F(\rho_{\alpha},\rho_{\beta})}\bigg],
\end{equation}
which holds for arbitrary lossy beam splitters and for any pair of coherent input states. 

Now we turn to the condition of coherent {\em perfect} absorption for the specific type of lossy beam splitters with $t=1/2$ and $r=-1/2$. These values for transmission and reflection amplitudes  give rise to maximum incoherent absorption of $1/2$ for a thin film (a beam-splitter) in a homogeneous background, meaning that for a single incident coherent state such beam splitters absorb half of the light since $|t|^{2} + |r|^{2} = 1/2 $. For the two incident coherent states we end up with the  coefficient for coherent absorption
\begin{equation}\label{eq:coh_ab_c_fid}
\mathcal{A}_{\text{coh}}^{\mathcal{C, I}}=1+\frac{\ln{\sqrt{F(\rho_{\alpha}, \rho_{\beta})}}}{|\alpha|^2+|\beta|^2}.
\end{equation}
Here the natural logarithm is well defined, since the fidelity of two arbitrary coherent states always is greater than zero due to their non-orthogonality property~(\ref{eq:not_orthogonal}). As a check we find back $\mathcal{A}_{\text{coh}}^{\mathcal{C, I}} = 1/2$ for a single incident coherent state (take $\alpha \ne\beta=0$).  

Coherent perfect absorption according to Eq.~(\ref{eq:coh_ab_c_fid}) occurs when $F(\rho_{\alpha}, \rho_{\beta})=1$, i.e., when the incoming coherent states are indistinguishable. This is satisfied if and only if the coherent states have the same phases and amplitudes. These are the the same well-known requirements of CPA as in classical optics: the expression~\ref{eq:coh_ab_c1} for the coherent absorption reduces to that obtained through classical scattering amplitudes for $t=1/2$ and $r=-1/2$~\cite{zhang2014coherent}. Thus with the ``quasi-classical'' coherent states we  recover the classical condition for CPA, but this time explained in terms of quantum mechanical indistinguishability. 

By discussing CPA in quantum rather than classical optics, we replaced interference  by quantum interference. The latter is described by transition probabilities between quantum states, which in the case of pure states, as we have here, are given by the fidelity~\cite{bengtsson2017geometry}. It is intriguing that  by Eqs.~\ref{eq:coh_ab_c2} or~\ref{eq:coh_ab_c_fid} fidelities can be measured in terms of  absorption. Thus, practically, one may distinguish two quantum states through a dissipative process, e.g., by using an absorbing metamaterial with known properties. 

\section{Coherent absorption of squeezed coherent states}\label{sec:CPA_sq}
\noindent Prepared by the theory and results for coherent states of Sec.~\ref{sec:results}, we will now investigate the effects of squeezing on the coherent absorption of light. Mathematically, a squeezed coherent state $|\alpha,\zeta\rangle$ is obtained by the action of a squeeze operator on a coherent state~\cite{scully1999quantum}, for example for mode~1
\begin{equation}\label{eq:sqeezed_state}
|\alpha_{1},\zeta_{1}\rangle_{1}=\hat{S}_{1}(\zeta_{1})|\alpha_{1}\rangle = \hat{S}_{1}(\zeta_{1})\hat{D}_{1}(\alpha_{1})|\rangle.
\end{equation} 
Here the squeeze operator is defined as
\begin{equation}
\hat{S}_{1}(\zeta_{1}) \equiv e^{\frac{1}{2}\zeta_{1}^{*}\hat{a}_{1}^{2}-\frac{1}{2}\zeta_{1}\hat{a}_{1}^{\dagger 2}},
\end{equation}
in terms of the mode creation and annihilation operators $a_{1}^{\dag}$ and $a_{1}$.
The degree of squeezing is determined by the complex coefficient $\zeta_{1}=\xi_{1}\exp{i\phi_{1}}$. Here, $\xi_{1}$ is called the squeezing parameter, while the angle $\phi_{1}$ quantifies the amount of rotation of the field quadratures in the corresponding quantum optical phase space.   

Let us now assume two  squeezed coherent states $|\alpha,\zeta_1\rangle_{\text{1}}$ and $|\beta,\zeta_2\rangle_{\text{2}}$ as the input states of a general  lossy beam splitter. The input states are then characterized by in total four complex parameters: the coherence parameters  $\alpha=|\alpha|\exp{(i\theta_1)}$ and $\beta=|\beta|\exp{(i\theta_2)}$, and the squeezing parameters $\zeta_1=\xi_1\exp{(i\phi_1)}$ and $\zeta_2=\xi_2\exp{(i\phi_2)}$. The total input state  has the form $|\psi\rangle_{\text{in}}=|\alpha,\zeta_1\rangle_{\text{1}}\otimes|\beta,\zeta_2\rangle_{\text{2}}\otimes|\rangle_{\text{BS}}$.
\subsection{Perfect coherent absorption of coherence}\label{Sec:CPA_coherence_squeezed}
\noindent The expected values of the input operators with respect to the state $|\psi\rangle_{\text{in}}$ read
\begin{subequations} 
\begin{eqnarray}
\langle\hat{a}_{1}\rangle&=&|\alpha|(e^{i\theta_1}\cosh{(\xi_1)}-e^{-i\theta_1}e^{i\phi_1}\sinh{(\xi_1)}),\\
\langle\hat{a}_{2}\rangle&=&|\beta|(e^{i\theta_2}\cosh{(\xi_2)}-e^{-i\theta_2}e^{i\phi_2}\sinh{(\xi_2)}).
\end{eqnarray}
\end{subequations}
For the input coherence~(\ref{eq:input_coherence}),i.e., the coherent content within squeezed states introduced by the Glauber's degree of freedom, it then follows that 
\begin{equation}\label{eq:c_inq}
\mathcal{C}_{\text{in}}=\gamma_1^2|\alpha|^2+\gamma_2^2|\beta|^2,
\end{equation}
where $\gamma_1^2=\cosh{(2\xi_1)}-\cos{(\eta_1)}\sinh{(2\xi_1)}$, $\gamma_2^2=\cosh{(2\xi_2)}-\cos{(\eta_2)}\sinh{(2\xi_2)}$ with $\eta_1 \equiv 2\theta_1-\phi_1$ and $\eta_2 \equiv 2\theta_2-\phi_2$. Similarly, by using the input-output relations~(\ref{eq:in_out}), we obtain the output coherence 
\begin{equation}\label{eq:c_outq}
\mathcal{C}_{\text{out}}=(|t|^2+|r|^2)C_{\text{in}}+\Gamma(tr^{*}+rt^{*}),
\end{equation}
where
\begin{eqnarray}\label{eq:gamma}
\nonumber\Gamma&=&\langle\hat{a}_{1}\rangle\langle\hat{a}_{2}\rangle^{*}+\langle\hat{a}_{2}\rangle\langle\hat{a}_{1}\rangle^{*}\\
\nonumber&=&2|\alpha||\beta|\big(\cos{(\theta)}\cosh{(\xi_1)}\cosh{(\xi_2)}\\
\nonumber&-&\cos{(\theta_1+\theta_2-\phi_2)}\cosh{(\xi_1)}\sinh{(\xi_2)}\\
\nonumber&-&\cos{(\theta_1+\theta_2-\phi_1)}\sinh{(\xi_1)}\cosh{(\xi_2)}\\
&+&\cos{(\theta+\phi)}\sinh{(\xi_1)}\sinh{(\xi_2)}\big),
\end{eqnarray}
and where $\theta = \theta_{2}-\theta_{1}$ and $\phi = \phi_{2}-\phi_{1}$.
Therefore, by Eq.~\ref{eq:coh_ab_c0}, the fraction of the coherence that gets lost is 
\begin{equation}\label{eq:sq_abs_coef}
\mathcal{A}_{\text{\rm sq}}^{\mathcal{C}}=1-\bigg(|t|^2+|r|^2+\frac{\Gamma(tr{^*}+rt^{*})}{\gamma_1^2|\alpha|^2+\gamma_2^2|\beta|^2}\bigg).
\end{equation}
In the following we will analyze whether it is possible that all coherence gets lost at the absorbing beam splitter, i.e. whether perfect absorption of coherent photons ($\mathcal{A}_{\text{\rm sq}}^{\mathcal{C}} = 1$) can be achieved with two incoming squeezed states. We will study the corresponding coherent absorption of intensities in Sec.~\ref{Sec:CPA_intensity_squeezed}. 

We study how squeezing affects the coherent  absorption of quantum coherence by exploring Eq.~\ref{eq:sq_abs_coef} for several parameter regimes.
Let us first assume two squeezed beams with unit coherent amplitudes, $|\alpha|=|\beta|=1$, with coherent phase angles $\theta_1=\theta_2=\delta$ and also equal squeezing angles $\phi_1=\phi_2=\phi$, such that $\delta\neq\phi$. So we take many parameters to be equal, but we do allow the squeezing amplitudes $\xi_1$ and $\xi_2$ to be different. It follows that
\begin{equation}\label{eq:sq_abs_coef2}
\mathcal{A}_{\text{\rm sq}}^{\mathcal{C}}=1-\big(|t|^2+|r|^2+\frac{\Omega_1}{\Omega_2}\big),
\end{equation}
where
\begin{eqnarray*}
\Omega_1&=&2[\cosh{(\xi^{\prime})}-\cos{(\epsilon)}\sinh{(\xi^{\prime})}](tr^{*}+rt^{*}),\\
\Omega_2&=&\cosh{(2\xi_1)}+\cosh{(2\xi_2)}\\
&-&\cos{(\epsilon)}[\sinh{(2\xi_1)}+\sinh{(2\xi_2)}],
\end{eqnarray*}
and we defined $\xi^{\prime}\equiv\xi_1+\xi_2$ and $\epsilon \equiv 2\delta-\phi$. We specify again $t=1/2$ and $r=-1/2$ and then find the condition for complete absorption of coherence to be $\Omega_{1}/\Omega_{2} = 1/2$, which is only satisfied if $\xi_1=\xi_2$. Therefore, for squeezed coherent states, equal coherent amplitudes and phases  are not sufficient criteria for perfect absorption of coherence. Equal degree of squeezing is an additional requirement. For coherent states this is trivially satisfied ($\xi_{1} = \xi_{2} = 0$). For squeezed states the additional requirement  is nontrivial, generalizing the requirement of indistinguishability  that we identified for coherent input states in Sec.~\ref{sec:CPA_fidelity}. 
\begin{figure}[t!]
	\centering
	\begin{center}
		\includegraphics[width=7cm]{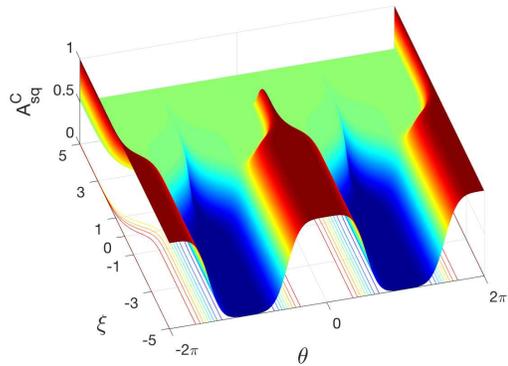}
	\end{center}
	\caption{Coefficient of coherent absorption $\mathcal{A}_{\text{\rm sq}}^{\mathcal{C}}(\xi,\theta)$ of Eq.~\ref{eq:coef_c_theta_r} upon variation of  parameters of the squeezed coherent input states, for a beam splitter with $t=1/2$ and $r=-1/2$. We vary the coherence angle $\theta = \theta_1$, and the squeezing parameter  $\xi = \xi_1=\xi_2$, while keeping $\theta_2=\phi_1=\phi_2=0$ fixed. Figure is valid for arbitrary equal coherence amplitudes $\alpha_{1} = \alpha_{2}$.
	}
	\label{fig:fig2}
\end{figure}

Further non-trivial effects of quantum squeezing on the absorption of quantum coherence can be revealed by considering the special case of two input states with
equal coherent amplitudes ($\alpha_{1} = \alpha_{2}$), a nonvanishing coherent phase difference ($\theta_1=\theta$ and $\theta_2=0$), 
equal squeezing amplitudes $\xi_1=\xi_2=\xi$ and vanishing squeezing phases ($\phi_1=\phi_2=0$). As before the absorbing beam splitter is characterized by $t=1/2$ and $r=-1/2$. 
We obtain
\begin{equation}\label{eq:coef_c_theta_r}
\mathcal{A}_{\rm sq}^{\mathcal{C}}=\frac{1}{2}+\frac{\cos{(\theta)}}{1+e^{2\xi}[\cosh{(2\xi)}-\cos{(2\theta)}\sinh{(2\xi)}]}.
\end{equation}
Thus, for all phase differences $\theta$, in the limit $\xi\rightarrow+\infty$ (amplitude squeezing), the coefficient of absorption of quantum coherence converges to the maximum incoherent absorption of $\mathcal{A}=1/2$. With little squeezing, up to $\xi\approx5$ say, (almost) perfect absorption of quantum coherence is recovered for $\theta$ equal to a multiple of $2\pi$. In the opposite limit $\xi\rightarrow-\infty$ (phase squeezing), the coefficient of absorption becomes $\mathcal{A}_{\rm sq}^{\mathcal{C}}=1/2+(\cos{(\theta)}/[1+(1/2)(1+\cos{(2\theta)})]$, i.e., we have a similar behavior as in the case of bare coherent states. An illustration of  these analytical results is depicted in Fig.~\ref{fig:fig2}. 

\begin{figure}[!t]
	\centering
	\begin{center}
		\subfigure[]{
			\label{fig:fig3a}
			\includegraphics[width=6cm]{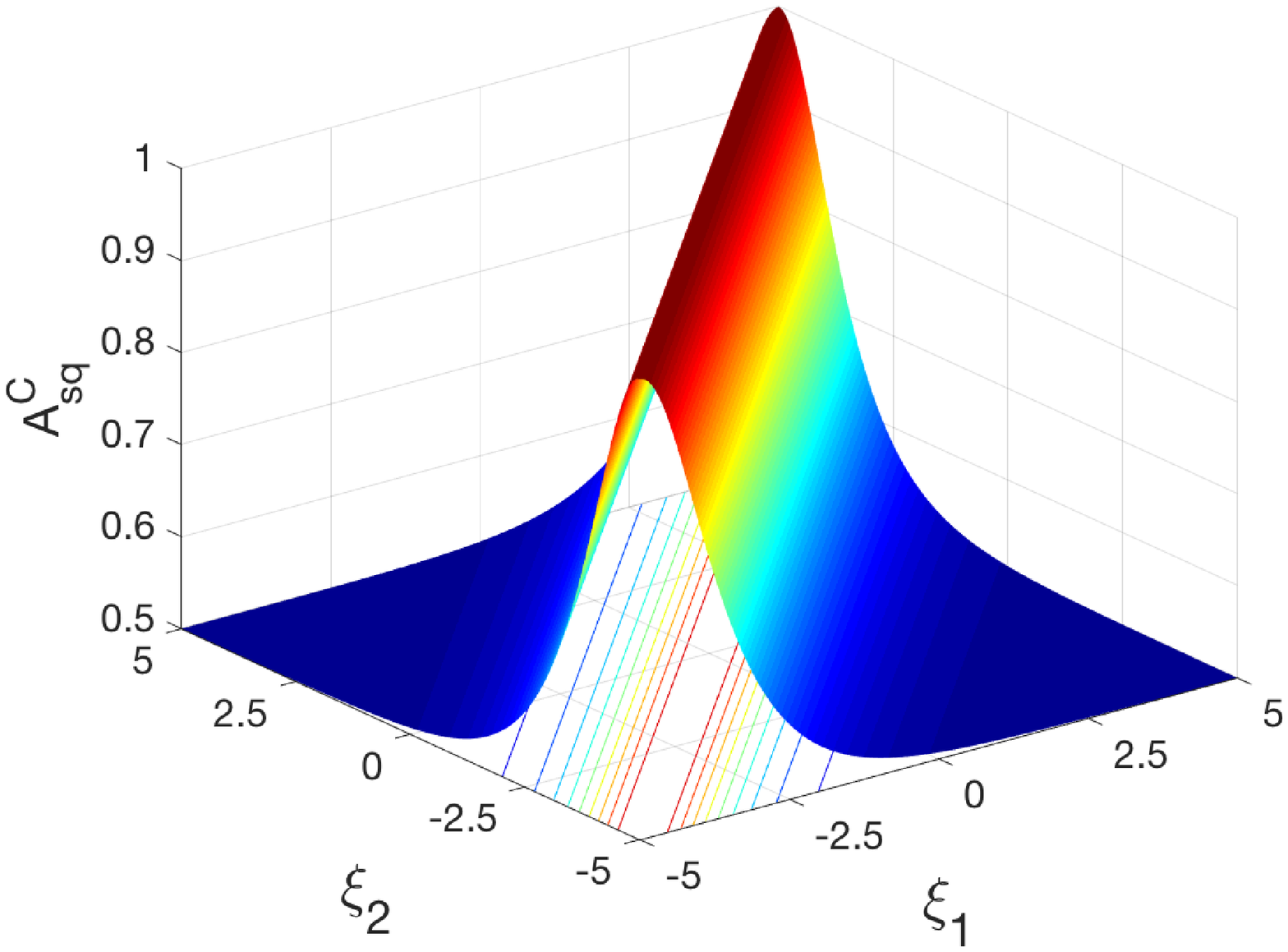}
		}\\
		\subfigure[]{
			\label{fig:fig3b}
			\includegraphics[width=5.5cm]{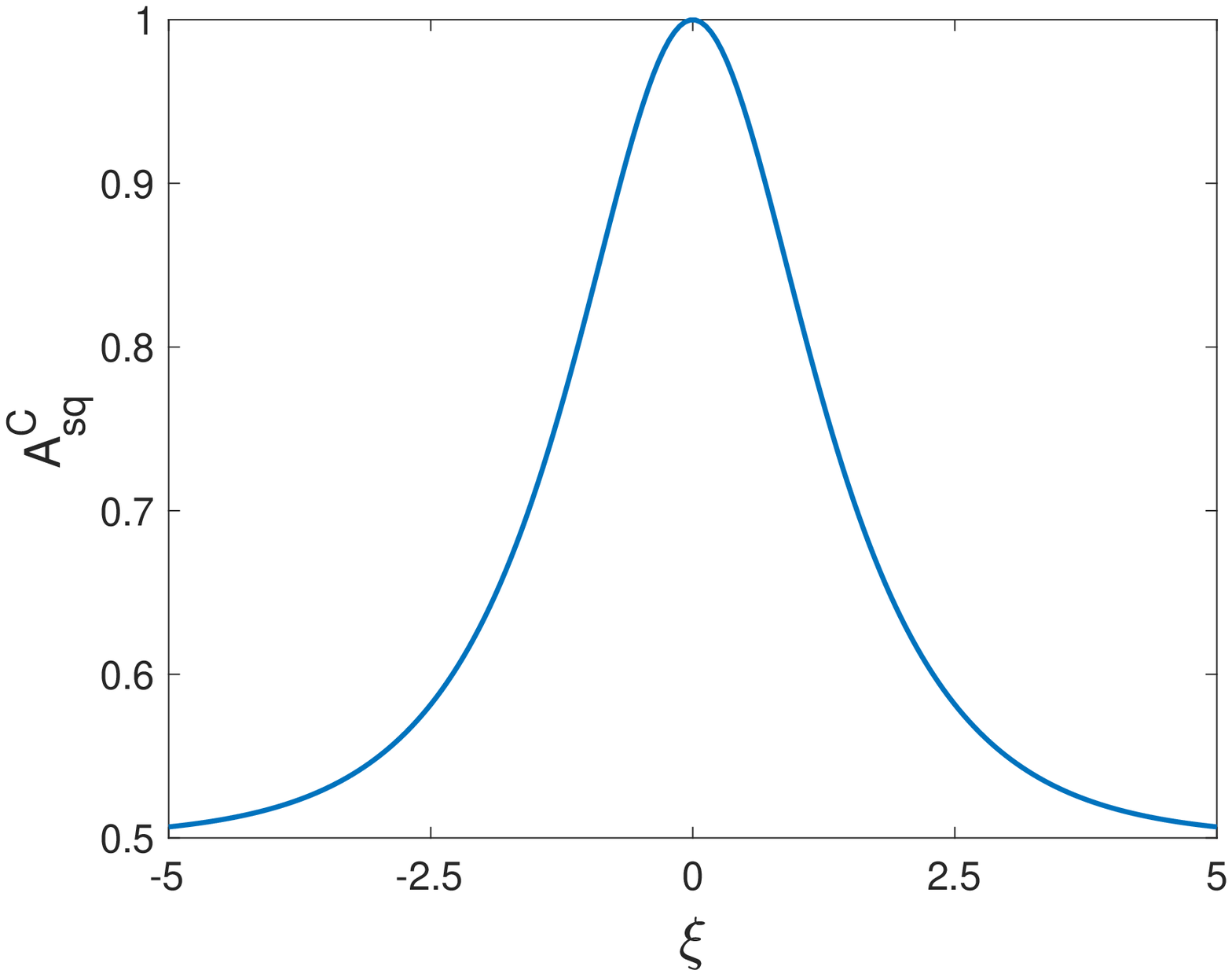}
		}
	\end{center}
	\caption{Variations in the coefficient of coherent absorption (a) $A_{\text{\rm sq}}^{\mathcal{C}}(\xi_1,\xi_2)$ of Eq.~\ref{eq:sq_abs_coef3} and (b) $A_{\text{\rm sq}}^{\mathcal{C}}(\xi_1=\xi,\xi_2=0)$ of Eq.~\ref{eq:sq_abs_coef5}. All other parameters are as explained in the main text.
	}
	\label{fig:fig3}
\end{figure}

While Eq.~\ref{eq:coef_c_theta_r} was found for different coherent phases but equal squeezing, as a final example to illustrate loss of coherence in the presence of squeezing, we revert the situation and take the coherence phases and amplitudes to be equal, but different squeezing amplitudes  (and again  $\delta=\phi=0$ and the same beam splitter with $t=1/2$ and $r=-1/2$). This gives
\begin{equation}\label{eq:sq_abs_coef3}
\mathcal{A}_{\rm sq}^{\mathcal{C}}=1-\frac{1}{2}\bigg(1-\frac{2e^{-\xi^{\prime}}}{e^{-2\xi_1}+e^{-2\xi_2}}\bigg).
\end{equation}
This formula is depicted in Fig.~\ref{fig:fig3} and implies that in the absence of phases, squeezing always works against absorption provided that $\xi_1\neq\xi_2$. Indeed, the perfect absorption of coherent photons occurs only if $\xi_1=\xi_2$. Furthermore, for equal squeezing we regain the symmetry such that in the limits $\xi\rightarrow\pm\infty$, the absorption saturates at its classical maximum of $0.5$. 
%
\subsection{Coherent absorption of intensity}\label{Sec:CPA_intensity_squeezed}
We showed that, for two coherent incident states, the fraction of quantum coherence lost is equal to that of intensity, i.e., $\mathcal{A}_{\text{coh}}^{\mathcal{C}}=\mathcal{A}_{\text{coh}}^{\mathcal{I}}$ or $\Delta\mathcal{C}-\Delta\mathcal{I}=0$. In the case of two squeezed incident states, the loss of intensity reads
\begin{equation}\label{eq:intensity_squeezed}
\Delta\mathcal{I}=\mathcal{I}_{\text{in}}\bigg(1-(|t|^2+|r|^2)-\frac{\Gamma(t^{*}r+r^{*}t)}{\mathcal{I}_{\text{in}}}\bigg),
\end{equation} 
where $\Gamma$ is given in Eq.~\ref{eq:gamma}. Hence, the coefficient of absorption of intensity becomes
\begin{eqnarray}\label{eq:abs_sq_int}
\nonumber \mathcal{A}_{\text{\rm sq}}^{\mathcal{I}}&=&1-(|t|^2+|r|^2)-\frac{\Gamma(t^{*}r+r^{*}t)}{\mathcal{I}_{\text{in}}},\\
&=&\mathcal{A}-\frac{\Gamma(t^{*}r+r^{*}t)}{\mathcal{I}_{\text{in}}},
\end{eqnarray}
which can be rewritten as
\begin{equation}\label{eq:sq_int_prime}
\mathcal{I}_{\text{in}}\mathcal{A}_{\text{\rm sq}}^{\mathcal{I}}=\mathcal{I}_{\text{in}}\mathcal{A}-\Gamma(t^{*}r+r^{*}t).
\end{equation}
Similarly, from Eq.~\ref{eq:sq_abs_coef}, we have
\begin{equation}\label{eq:sq_abs_prime}
\mathcal{C}_{\text{in}}\mathcal{A}_{\text{\rm sq}}^{\mathcal{C}}=\mathcal{C}_{\text{in}}\mathcal{A}-\Gamma(t^{*}r+r^{*}t).
\end{equation}
Eq.~\ref{eq:sq_int_prime} and~\ref{eq:sq_abs_prime} can be combined with $\mathcal{C}_{\text{in}}\mathcal{A}_{\text{\rm sq}}^{\mathcal{C}}=\Delta\mathcal{C}$ and $\mathcal{I}_{\text{in}}\mathcal{A}_{\text{\rm sq}}^{\mathcal{I}}=\Delta\mathcal{I}$ 
to give
\begin{equation}\label{eq:general_formula}
\Delta\mathcal{I}-\Delta\mathcal{C}=(\mathcal{I}_{\text{in}}-\mathcal{C}_{\text{in}})\mathcal{A}.
\end{equation}
Important about this identity is its generality: it relates coherence losses and intensity losses for arbitrary lossy beam splitters and arbitrary squeezed coherent input states. 
It is the generalization of the simple identity  $\Delta\mathcal{I}=\Delta\mathcal{C}$ that we obtained for  coherent states. 
If $\mathcal{A}_{\text{\rm sq}}^{\mathcal{I}},\mathcal{A}_{\text{\rm sq}}^{\mathcal{C}}\neq\mathcal{A}$, then Eq.~\ref{eq:general_formula} implies that one has equal quantum absorption coefficients $\mathcal{A}_{\text{\rm sq}}^{\mathcal{I}}=\mathcal{A}_{\text{\rm sq}}^{\mathcal{C}}$ if and only if the total input intensity is equal to the total input coherence, i.e. $\mathcal{C}_{\text{in}}=\mathcal{I}_{\text{in}}$. But if the latter are not equal, and the coherent part of the squeezed state is completely absorbed ($\Delta\mathcal{C} =C_{i}$), then it follows from Eq.~\ref{eq:general_formula} that $I_{\rm out} = (I_{\rm in}-C_{\rm in})(1-A)\ne 0$, in other words a quantum state with finite intensity survives  the coherent absorption process. We will analyze this output state in Sec.~\ref{sec:CPA_qsp}.

\begin{figure}[t!]
	\centering
	\begin{center}
		\includegraphics[width=7cm]{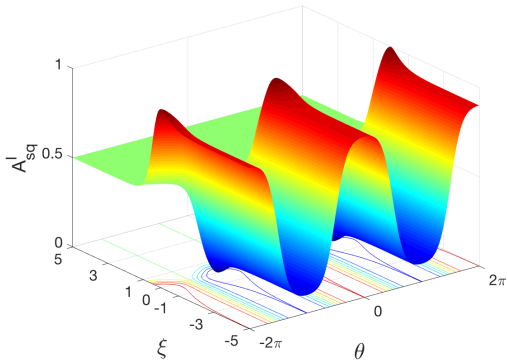}
	\end{center}
	\caption{Same as Fig.~\ref{fig:fig2} but for $\mathcal{A}_{\text{\rm sq}}^{\mathcal{I}}(\xi, \theta)  $ with $|\alpha|=|\beta|=1$. 
	}
	\label{fig:fig4}
\end{figure}

For a fair comparison of coherence and intensity absorption, we now consider the same special cases that we already investigated in our analysis of the coefficient of absorption of quantum coherence. First, we  choose $\theta_1=\theta$, $\xi_1=\xi_2=\xi$, $\theta_2=\phi_1=\phi_2=0$ and $t=1/2$ and $r=-1/2$ with equal coherent amplitudes. We obtain
\begin{equation}
\mathcal{A}_{\rm sq}^{\mathcal{I}}=\frac{1}{2}+\frac{(\cos{(\theta)}/2)e^{-2\xi}}{e^{-2\xi}+[\frac{1-\cos{(2\theta)}}{2}]\sinh{(2\xi)}+\frac{\sinh^2{(\xi)}}{|\alpha|^2}},
\end{equation}
which should be compared with Eq.~\ref{eq:coef_c_theta_r} that gives $\mathcal{A}_{\rm sq}^{\mathcal{C}}$ for the same input states. 
In the limit $\xi\rightarrow+\infty$, we have $\mathcal{A}_{\rm sq}^{\mathcal{I}}=1/2$ as before. Thus, in this limit all the quantum contributions due to quantum coherence and squeezing are lost and the corresponding absorption coefficients reduce to that of maximum incoherent one, i.e., $\mathcal{A}_{\rm sq}^{\mathcal{C}}=\mathcal{A}_{\rm sq}^{\mathcal{I}}=\mathcal{A}$. In the opposite limit of $\xi\rightarrow-\infty$, we obtain $\mathcal{A}_{\rm sq}^{\mathcal{I}}=(1/2)+2\cos{(\theta)}/[3+\cos{(2\theta)}+(1/|\alpha|^2)]<1$ for all $\alpha\in\mathbb{C}$, thus characteristically different than absorption of quantum coherence but complying with our general relation~\ref{eq:general_formula}. Fig.~\ref{fig:fig4} illustrates that perfect absorption of intensity is possible if and only if there is no squeezing. The discrepancy between absorption of coherence in Fig.~\ref{fig:fig2} and of intensity in Fig.~\ref{fig:fig4} is evident. The quantum coefficient $\mathcal{A}_{\rm sq}^{\mathcal{I}}$  saturates to its classical incoherent value of $1/2$ faster than $\mathcal{A}_{\rm sq}^{\mathcal{C}}$ for $\xi>0$. In the opposite regime where $\xi<0$, we have coherent oscillations similar to the case of absorption of quantum coherence, though always in the interval $[0,1)$.

A crucial difference between the coefficients $\mathcal{A}_{\rm sq}^{\mathcal{C}}$ and $\mathcal{A}_{\rm sq}^{\mathcal{I}}$ is that the latter depends explicitly on the mean number of photons in the initial states. This leads to the breaking of the parity symmetry such that $\mathcal{A}_{\rm sq}^{I}(\xi,\theta=0,\delta=0)\neq\mathcal{A}_{\rm sq}^{I}(-\xi,\theta=0,\delta=0)$. To clarify this, let $\xi_2=0$ and $\xi_1=\xi$, so that the first beam is prepared in a squeezed state, while the second one is in a coherent state with equal amplitudes $|\alpha|$. By setting all optical phases to zero, the coefficient of absorption of coherent degree of freedom of squeezed states~\ref{eq:sq_abs_coef3} is found to be
\begin{subequations}
\begin{equation}\label{eq:sq_abs_coef5}
 \mathcal{A}_{\rm sq}^{\mathcal{C}}=\frac{1}{2}+\frac{e^{-\xi}}{1+e^{-2\xi}}
=\frac{1+\cosh\xi}{2\cosh\xi},
\end{equation}
while the intensity absorption coefficient  becomes
\begin{equation}\label{eq:ab_r_alfa_i}
\mathcal{A}_{\rm sq}^{\mathcal{I}}=\frac{1}{2}+\frac{e^{-\xi}}{1+e^{-2\xi}+\frac{\sinh^2{(\xi)}}{|\alpha|^2}}.
\end{equation}
\end{subequations}
%
\begin{figure}[t!]
	\centering
	\begin{center}
		\includegraphics[width=6.5cm]{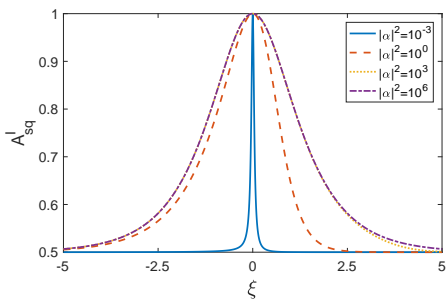}
	\end{center}
	\caption{Variations in the coefficient of coherent absorption of intensity $A_{\text{\rm sq}}^{\mathcal{I}}(\xi_1=\xi,\xi_2=0)$ of Eq.~\ref{eq:ab_r_alfa_i} for $|\alpha|^2=10^{-3}$ (solid-blue), $|\alpha|^2=10^{0}$ (dashed-orange), $|\alpha|^2=10^{3}$ (dotted-yellow) and $|\alpha|^2=10^{6}$ (dashed-dotted-purple). All the other parameters are as explained in the main text.
	}
	\label{fig:fig5}
\end{figure}
%
Hence, in this regime, the coefficient $\mathcal{A}_{\rm sq}^{\mathcal{C}}$ is an even function and thus totally symmetric in the squeezing parameter $\xi$,  see Fig.~\ref{fig:fig3b}. The symmetry-breaking term in the expression for $\mathcal{A}_{\rm sq}^{\mathcal{I}}$ is now easily recognized as the ratio $R=\sinh^2{(\xi)}/|\alpha|^2$ of the mean photon contributions of independent squeezing and coherent degrees of freedoms to the total intensity of the squeezed coherent state. The maximally asymmetric and symmetric regimes are then identified by $R\gg1$ and $R\ll1$, respectively. In Fig.~\ref{fig:fig5}, we plot the coefficient of coherent absorption of intensity $A_{\text{\rm sq}}^{\mathcal{I}}(\xi_1=\xi,\xi_2=0)$ of Eq.~\ref{eq:ab_r_alfa_i}, as a function of the squeezing parameter $\xi$, for a set of coherent amplitudes $|\alpha|^2\in\{10^{-3},10^{0},10^{3},10^{6}\}$. The figure illustrates clearly that intensity absorption in general is not symmetric in $\xi$. To restore the symmetry in the considered regime, the minimum number of coherent photons is found to be of order $10^{6}$.
\section{Continuous-variable quantum state preparation with CPA}\label{sec:CPA_qsp}
We would like to know the quantum states of light produced at the output of a beam splitter that exhibits CPA. For equal coherent states $|\alpha\rangle_{1}$ and $|\alpha\rangle_{2}$ as input, we found in Sec.~\ref{sec:2B} that there is zero intensity in the output, meaning that the output state for the two optical output modes has to be the vacuum state. The remarkable robustness of this way of producing the vacuum as output is that the same output state is produced whatever  the coherence amplitude $|\alpha|$ of the input state. 

Quantum state preparation becomes even more interesting for squeezed coherent input states, which we write in terms of squeezing and displacement operators as
\begin{eqnarray}\label{eq:in_state_ops}
\nonumber|\psi\rangle_{\text{in}}&=&|\alpha,\zeta_1\rangle_{1}\otimes|\beta,\zeta_2\rangle_{2}\otimes|\rangle_{\text{BS}}\\
\nonumber&=&\hat{S}_{2}(\zeta_2)\hat{D}_{2}(\beta)\hat{S}_{1}(\zeta_1)\hat{D}_{1}(\alpha)|\rangle\otimes|\rangle_{\text{BS}}\\
\nonumber&=&\hat{S}_{1}(\zeta_1)\hat{S}_{2}(\zeta_2)\hat{D}_{1}(\alpha)\hat{D}_{2}(\beta)|\rangle\otimes|\rangle_{\text{BS}}\\
\nonumber&=&
e^{\frac{1}{2}(\zeta_2^{*}\hat{a}_{2}^2+\zeta_1^{*}\hat{a}_{1}^2)-\frac{1}{2}(\zeta_2\hat{a}_{2}^{\dagger 2}+\zeta_1\hat{a}_{1}^{\dagger 2})}
\\
&\times&
\hat{D}_{1}(\alpha)\hat{D}_{2}(\beta) |\rangle\otimes|\rangle_{\text{BS}},
\end{eqnarray} 
where in the third and fourth equalities we used that $\hat{a}_{1}$ and $\hat{a}_{2}^{\dag}$ commute. In the previous section we saw that for equal input amplitudes and squeezing, and for $t = -r = 1/2$, all coherence can be coherently absorbed but some output intensity will always remain. We will now use the input state~(\ref{eq:in_state_ops}), specify $\beta = \alpha$ and $\zeta_{2}=\zeta_{1}=\zeta$, and then determine the output state for this specific case. 

In Eq.~\ref{eq:in_out}, output operators were defined in terms of input operators. We need to invert this, writing the input operators in terms of the output operators. Thereby we can obtain the sought output state by writing the input state in terms of the output operators. 
We will use the known quantum optical input-output theory for absorbing beam splitters~\cite{PhysRevA.52.4823,PhysRevLett.77.1739,PhysRevA.57.2134,Knoell:1999a}, in particular Ref.~\cite{Knoell:1999a}, and identify what is special about quantum state transformation by absorbing beam splitters that exhibit CPA.  
Following Ref.~\cite{Knoell:1999a} we write the input-output operator relations Eq.~\ref{eq:in_out} in matrix notation as \begin{equation}\label{Eq:outputoperatorsasinput}
\hat{\bf b}={\bf T}\hat{\bf a}+{\bf A}\hat{\bf g}.
\end{equation}
Here ${\bf T}$ is the $2\times 2$ transmission matrix. The Langevin noise of the absorbing beam splitter is accounted for by  linear combinations of bosonic device input operators $\hat g_{1}$ and $\hat g_{2}$ that together form the vector $\hat{\bf g}$. The corresponding linear coefficients form the $2\times 2$ absorption matrix ${\bf A}$. Besides optical output operators $\hat{b}_{1,2}$, there are device output operators $\hat{h}_{1,2}$. Also the latter pair can be written as a linear combination of all four input operators. The $4\times 4$ matrix  that relates all four output operators in terms of the four  input operators is restricted by the requirement that  output operators  satisfy standard bosonic commutation relations and are canonically independent. This restricts ${\bf A}$ once ${\bf T}$ is given, for example.
 
The formalism of Ref.~\cite{Knoell:1999a} simplifies particularly for the CPA beam splitter because ${\bf T}$ is a real symmetric matrix in this special case, and the absorption matrix ${\bf A}$ is then also easily found:
\begin{eqnarray}
 {\bf T}_{\text{cpa}}  =  \left(
 \begin{array}{cc}
 t & r \\ r & t 
 \end{array}
 \right) & = & \frac{1}{2}\left(\begin{array}{cc}
 1 & -1 \\ -1 & 1 
 \end{array}
 \right) = \frac{1}{2} \left( \openone - \sigma_{x}\right), \nonumber \\
 {\bf A}_{\text{cpa}} \qquad\qquad \quad & = & \frac{1}{2}\left(\begin{array}{cc}
 1 & 1 \\ 1 & 1 
 \end{array}
 \right) = \frac{1}{2} \left( \openone + \sigma_{x}\right), 
 \end{eqnarray}
both  in terms of the $2\times 2$ unit matrix $\openone$ and the Pauli matrix $\sigma_{x}$. For the CPA beam splitter, the inverse relationship of Eq.~\ref{Eq:outputoperatorsasinput} becomes
\begin{equation}\label{Eq:inputoperatorsasoutputCPA}
\hat{\bf a}={\bf T}_{\text{cpa}}\,\hat{\bf b}-{\bf A}_{\text{cpa}}\,\hat{\bf h}.
\end{equation} 
Now we can use these relations to write $\hat{a}_{1}$ and $\hat{a}_{2}$ in the input state~(\ref{eq:in_state_ops}) in terms of the four output operators $\hat{b}_{1,2}$ and $\hat{h}_{1,2}$. We thereby obtain as one of our main results the output state 
\begin{eqnarray}\label{eq:output_state}
\nonumber|\psi\rangle_{\text{out}}&=&
e^{\frac{1}{4}\zeta^{*}(\hat{b}_{1}-\hat{b}_{2})^2-\frac{1}{4}\zeta(\hat{b}^{\dagger}_{1}-\hat{b}^{\dagger}_{2})^2}|\rangle
\\
&\otimes&
e^{\frac{1}{4}\zeta^{*}(\hat{h}_{1}+\hat{h}_{2})^2-\frac{1}{4}\zeta(\hat{h}^{\dagger}_{1}+\hat{h}^{\dagger}_{2})^2}|\alpha,\alpha\rangle_{\text{BS}},
\end{eqnarray} 
which is  remarkable for several reasons. First, just like the input state it is a direct-product state of optical output states and beam splitter device states. In other words, the optical output state is {\em not entangled} with the absorbing beam splitter that was used to produce it. Tracing out the beam splitter's internal degrees of freedom therefore leaves the output state of light in a pure state, rather than the usual mixed state that requires a density-matrix description. This remarkable outcome is the main reason why the CPA beam splitter, despite being lossy, can become a useful component in continuous-variable quantum state engineering.  

The second remarkable property of the state is the perfect coherent absorption: all coherence of the input state $|\psi\rangle_{\text{in}}=|\alpha,\zeta \rangle_{1}\otimes|\alpha,\zeta\rangle_{2}\otimes|\rangle_{\text{BS}}$ resided in the optical channels and ends up in the material modes of the beam splitter. There are no coherent photons, in Glauber's sense, in the optical output state, i.e. it does not depend on the coherence amplitude $\alpha$ at all. This explains that we found $A_{\text{\rm sq}}^{\mathcal{C}}=1$ in Sec.~\ref{Sec:CPA_coherence_squeezed}.

As a special case and check of our results, for vanishing squeezing we indeed find standard CPA behavior: for the 2-mode coherent input state  $|\psi\rangle_{\text{in}}=|\alpha \rangle_{1}\otimes|\alpha\rangle_{2}\otimes|\rangle_{\text{BS}}$ we find from Eq.~\ref{eq:output_state} the  corresponding output state $|\psi\rangle_{\text{out}}=|\rangle\otimes|\alpha,\alpha\rangle_{\text{BS}}$. This  is a direct product of indeed the optical vacuum state and coherent states for the device modes of the beam splitter. So for coherent states the coherent absorption is indeed perfect, no photons leave the CPA beam splitter and $A_{\text{\rm sq}}^{\mathcal{I}}=1$ . 

Returning to the general case of squeezed coherent input, the optical output state $e^{\frac{1}{4}\zeta^{*}(\hat{b}_{1}-\hat{b}_{2})^2-\frac{1}{4}\zeta(\hat{b}^{\dagger}_{1}-\hat{b}^{\dagger}_{2})^2}|\rangle$ in Eq.~\ref{eq:output_state}
 is a  one- and two-mode combined squeezed vacuum state. The one-mode squeezing corresponds to quadratic operators in the exponent  such as $\hat{b}_{2}^2$ , and the two-mode squeezing to the products of different operators such as $\hat{b}_{1}\hat{b}_{2}$.Being  generalizations of the generic squeezed vacuum, the optical output states can be used for the implementations of quantum teleportation~\cite{milburn1999quantum}, quantum metrology~\cite{anisimov2010quantum}, quantum dense coding~\cite{ban1999quantum}, quantum dialogues~\cite{zhou2017new} and electromagnetically induced transparency protocols~\cite{akamatsu2004electromagnetically}. Squeezed vacuum states have non-vanishing intensities, and since a beam splitter under CPA conditions emits squeezed vacuum states of light, coherent perfect absorption of intensity is not possible, and  $A_{\text{\rm sq}}^{\mathcal{I}}<1$ for non-vanishing squeezing.  This optical output state is independent of the coherence amplitude $\alpha$ of the incident squeezed coherent states. This quantum property explains why the coherent absorption coefficient $A_{\text{\rm sq}}^{\mathcal{I}}$ in Eq.~\ref{eq:ab_r_alfa_i} became dependent on the input intensity via $|\alpha|^2$, while  such a nonlinear dependence is absent for 
$A_{\text{\rm sq}}^{\mathcal{C}}$ in Eq.~\ref{eq:sq_abs_coef5}.

These one- and two-mode combined squeezed vacuum states have been studied before, albeit in a different setting~\cite{hong1990squeezed}. The exact form that emerges here was first proposed by Abdalla~\cite{abdalla1992statistical1,abdalla1992statistical} and later studied by Yeoman and Barnett~\cite{yeoman1993two}. In the latter contribution, it was identified that these states could be generated by superposing two identical (equally squeezed) single-mode squeezed vacuum states via a $50/50$ \textit{ideal} beam splitter. It was found that $|t|^2=|r|^2$ should hold to produce such states on a beam splitter, a condition that also the lossy CPA beam splitter satisfies. 
  
  So while the ideal beam splitter requires squeezed vacuum input states, the CPA beam splitter can take {\em any} pair of identical squeezed coherent states to distill~\cite{heersink2006distillation} a 2-mode entangled squeezed vacuum state out of it. Moreover, squeezing takes place via an absorption process resulting in beam-splitter internal modes that end up in  one- and two-mode combined squeezed coherent states. We find that half of the squeezing is absorbed into internal modes of the beam splitter. This leaves the other 50 percent of the squeezing for the optical output modes, in accordance with a spectral analysis performed for the special case of incoming  squeezed vacuum states~\cite{Huang:2014a}. 
The distillation of squeezed vacuum states that we propose is not possible with non-absorbing beam splitters. Thus, our results generalize the previous ones and propose a new engineering procedure to produce pure quantum states via perfect absorption of quantum coherence. 
\section{Conclusions}\label{sec:conc}
In conclusion, we investigated the coherent absorption of light when two squeezed coherent beams are superposed on an absorbing beam splitter. We first reconsidered the generic case of two incoming bare coherent states and distinguished two types of absorption,  namely of quantum coherence introduced to the system by the complex amplitude $\alpha$ and of intensity. We showed that the corresponding absorption coefficients are identical for the case of bare coherent state inputs and  can be written in terms of quantum fidelity, suggesting a general condition of indistinguishability of input states for CPA to occur that holds for squeezed coherent states as well.  

In the case of squeezed coherent beams, the coherent degree of freedom is completely absorbed, provided that the CPA conditions hold. By Eq.~\ref{eq:general_formula} we provided a general argument that the  input intensity will not be fully absorbed in the presence of quantum squeezing. More specifically, we  revealed that an entangled  squeezed vacuum state is produced  at the output, leaving the absorber in a squeezed coherent state. In some cases both states might be reused as quantum resources~\cite{winter2016operational,brandao2013resource}.

We propose to test and use the lossy CPA gate as a new tool for quantum state preparation. Since quite remarkably the CPA gate produces a direct-product state of an optical output state and an internal beam-splitter state~(see Eq.~\ref{eq:output_state}), it does not suffer from the usual disadvantage of lossy optical components that they become entangled with optical fields, producing mixed reduced quantum states for the light fields. Instead, the optical output states of the CPA gate are pure quantum states. Yet the action of the gate crucially depends on the CPA beam splitter being lossy: all coherence is absorbed. 

It is interesting to compare the CPA gate with the usual practical implementation of  ``phase-space displacement'' by which a squeezed vacuum state and a strong coherent state are mixed on a low-reflectivity non-lossy beam splitter~\cite{Lvovsky:2016a}, resulting in a squeezed coherent output state. Our CPA gate does more or less the reverse, separating squeezing from coherence, but the crucial difference is that it does so for {\em arbitrary} (but equal) input coherence amplitudes. This arbitrariness constitutes a useful robustness of this gate. In particular, the CPA gate  would work in a small-signal regime where saturation effects in absorption can safely be neglected.  

Our proposal of the  CPA quantum gate is part of an interesting wider trend to engineer quantum dissipation and to use it for quantum state preparation and other quantum operations, see for example Refs.~\cite{Kastoryano:2011a,Krauter:2011a,Bhaktavatsala:2014a,Kienzler:2015a,Metelmann:2015a,Morigi:2015a,Bhaktavatsala:2017a,Vest:2017a}. Also for our CPA gate for continuous-variable quantum state preparation, loss is a resource to obtain new functionality: the CPA gate prepares its pure quantum states both despite being lossy and because it is lossy.

\section*{Acknowledgments}
We thank E.C.~Andr{\'e}, N.~Stenger and J.R. Maack for useful discussions. A.\"{U}.C.H.~acknowledges the support from the Villum Foundation through
a postdoctoral Block Stipend. The Center for Nanostructured
Graphene is sponsored by the Danish National Research Foundation, Project DNRF103.
\bibliographystyle{apsrev4-1}
\bibliography{refs}
\end{document}